# A simple theory of the Invar effect in iron-nickel alloys

François Liot[1] & Christopher A. Hooley[2]

[1]*Department of Physics, Chemistry, and Biology (IFM), Linköping University, SE-581 83 Linköping, Sweden;* [2]*Scottish Universities Physics Alliance (SUPA), School of Physics and Astronomy, University of St Andrews, North Haugh, St Andrews, Fife KY16 9SS, U.K.*

**Certain alloys of iron and nickel (so-called 'Invar' alloys) exhibit almost no thermal expansion over a wide range of temperature[1–3]. It is clear that this is the result of an anomalous contraction upon heating which counteracts the normal thermal expansion arising from the anharmonicity of lattice vibrations. This anomalous contraction seems to be related to the alloys' magnetic properties, since the effect vanishes at a temperature close to the Curie temperature. However, despite many years of intensive research, a widely accepted microscopic theory of the Invar effect in face-centered-cubic Fe-Ni alloys is still lacking. Here we present a simple theory of the Invar effect in these alloys based on Ising magnetism, *ab initio* total energy calculations, and the Debye-Grüneisen model[4]. We show that this theory accurately reproduces several well known properties of these materials, including Guillaume's famous plot[1] of the thermal expansion coefficient as a function of the concentration of nickel. Within the same framework, we are able to account in a straightforward way for experimentally observed deviations from Vegard's law[2, 3, 5, 6]. Our approach supports the idea that the lattice constant is governed by a few parameters, including the fraction of iron-iron nearest-neighbour pairs.**

It is over eighty years since Guillaume received a Nobel prize for his discovery of the Invar effect. In 1897, he had observed that certain alloys of iron and nickel exhibit almost zero thermal expansion over a wide range of temperature[1]. The decades since then have seen many more precise measurements on the Fe-Ni series[2, 3], as well as the observation of a similar effect in other systems such as Fe-Pt[3] and Ni-Mn[3]. Some materials have also been shown to exhibit the opposite: an anomalously large thermal expansion, referred to as the 'anti-Invar' effect[6].

Naturally there has been a great deal of theoretical work on the origin of the Invar effect in face-centred-cubic (fcc) Fe-Ni alloys. The relevant measured quantity is the linear thermal expansion coefficient, $(dL/dT)/L$, which we denote by $\alpha$. This may be decomposed as $\alpha = \alpha_{latt} + \alpha_{anom}$, where $\alpha_{latt}$ is the contribution from the lattice vibrations. There appears to be an almost universal consensus that the anomalous term $\alpha_{anom}$ is related to the magnetic properties of the systems in question.

On the issue of which magnetic models are appropriate, however, opinions differ. One strand in the literature favours a so-called 2γ-state model, where the iron atoms in the alloy switch between two states with different magnetic moments (and thus different volumes) as the temperature is raised[3]. This approach, however, appears to be incompatible with the results of Mössbauer[7] and neutron[8] experiments. A second approach is based on weak itinerant ferromagnetism[3]; this, in turn, seems incompatible with the experimental observation[9] that substantial local magnetic moments persist in the paramagnetic phase of $Fe_{65}Ni_{35}$.

Models based on local moments[3, 10–14] have also been suggested, a point of view that is supported by certain Mössbauer[15] experiments. Within this class of models, there has been some emphasis recently[12] on the importance of non-collinearity of the magnetic moments on the iron sites, though experiments[16] undertaken to detect such non-collinearity do not seem to find it. Other authors, by contrast, conclude[11] that an Ising ('up or down') model with properly chosen exchange constants is adequate to the problem – a view supported by our results, as we shall demonstrate below.

An important step towards a quantitative prediction of the thermal expansion of Invar alloys has recently been made using the partially disordered local moment (PDLM) model[17]. This model can be used to calculate *ab initio* total energies of an $A_{1-x}B_x$ alloy in collinear magnetic states, given the fraction of A atoms whose spins are 'up', $x_{AU}$, and the fraction of B atoms whose spins are 'up', $x_{BU}$. A PDLM state for $x_{AU}=x_{BU}=1$ is fully magnetised, whereas a PDLM state for $x_{AU}=x_{BU}=0.5$ is completely disordered. By applying this formalism, the Invar effect in disordered Fe-Pt alloys has been found to originate from thermal magnetic disorder.

In this work, we present a simple theory of the relationship between magnetism and thermal expansion in fcc Fe-Ni alloys. X-ray diffraction measurements[18, 19] have shown that fcc Fe-Ni alloys are chemically disordered. We treat them as random alloys, calculating the Wigner-Seitz radius by minimising the free energy within the Debye-Grüneisen model at the temperature of interest, $T$. For this purpose, we use a partially disordered local moment state exhibiting similar magnetic properties to those of the random alloy at $T$.

The following general procedure is used to determine the average lattice constant at temperature $T$. (See the Methods section for details of the calculations.)

1. We calculate the following characteristics of the magnetic structure at $T$: the magnetization $M(T)$; the fraction of nearest-neighbour iron-iron pairs whose spins are anti-parallel, $X_{FFAP}(T)$; the analogous quantity for iron-nickel pairs, $X_{FNAP}(T)$; and the same for nickel-nickel pairs, $X_{NNAP}(T)$. As discussed above, an Ising model seems to be sufficient for this step of the calculation; see the Methods section for further details.

2. We use in our *ab initio* code a PDLM state that reproduces the values of $X_{FFAP}$, $X_{FNAP}$, $X_{NNAP}$, and $M$ found in part 1. In our Ising approach, this state is labelled by the fraction of nickel spins that are up ($x_{NU}$) and the fraction of iron spins that are up ($x_{FU}$). We thus obtain the total energy for that PDLM state as a function of the lattice constant.

3. We fit a Morse function to the results of part 2, using a least-squares procedure. The parameters of the fit give us the equilibrium lattice constant, the bulk modulus, and the Grüneisen constant[12] for the values of $x_{FU}$ and $x_{NU}$ in question.

4. Using the Debye-Grüneisen model, we calculate the free energy as a function of the lattice constant. This is done by adding the total energy obtained in part 2 to the free energy of the vibrating lattice estimated from the three quantities obtained in part 3[4]. Minimising this free energy allows us to calculate the average lattice constant at temperature $T$.

A useful way to think of the above procedure is shown in Fig. 1. Imagine that the magnetic configuration were fixed. Then the material would show only "normal" thermal expansion. Let us call the corresponding thermal expansion curve $a(x_{FU}, x_{NU}, T)$; the curve for $x$=0.35, $x_{FU}$=0.90, and $x_{NU}$=1 is the uppermost dashed curve in Fig. 1. In reality, however, raising the temperature causes the material to demagnetise, and the values of $X_{FFAP}$, $X_{FNAP}$ and $X_{NNAP}$ change accordingly. One may say that the system 'hops' from the curve $a(x_{FU}, x_{NU}, T)$ to the curve $a(x'_{FU}, x'_{NU}, T')$, resulting in a lattice spacing given by the curve $a(x_{FU}(T), x_{NU}(T), T)$. This is shown as the solid line in Fig. 1. In the case depicted, each 'hop' is to a curve lower than the last, cancelling the upward trend of each individual curve: this is the essence of the Invar effect.

Having thus outlined the general procedure to be followed, we now present the details, with some commentary. In Fig. 2a, we plot the fraction of nearest-neighbour Fe-Fe pairs that are anti-parallel, $X_{FFAP}$, as a function of temperature, according to the Ising model we use; these results are shown for several different overall concentrations of Ni in the alloy. Fig. 2b displays the equilibrium lattice constant of $Fe_{1-x}Ni_x$ in the PDLM state with the appropriate $x_{FU}$ and $x_{NU}$ for $x$=0.30, 0.35, 0.45, 0.55, and 0.80, versus $X_{FFAP}$. Note that the lattice constant is strongly negatively correlated with the number of anti-parallel iron-iron pairs. This reduction of the lattice constant as the number of iron-iron ↑↓ pairs is increased is the *central physical mechanism* of the Invar effect. It is interesting to note that a clear shift of the equilibrium volume towards lower values with increasing the number of iron-iron ↑↓ pairs can also be observed in Invar Fe-Pt alloys[17].

We now employ the above-described procedure (points 3 and 4) to calculate the lattice constant, $a$, as a function of temperature. The results are shown in Fig. 3a, exhibiting a clear Invar effect for certain concentrations $x$. From this, we may easily extract the linear thermal expansion coefficient at room temperature, which is plotted in Fig. 3b. Note the quantitative agreement with experiment, both for concentrations that exhibit the Invar effect and for those that do not.

There is one question, however, that our *ab initio* calculations so far do not answer: namely, why is it that an increase in the number of nearest-neighbour iron-iron ↑↓ pairs, $X_{FFAP}$, causes a reduction in the lattice spacing? To address this point, we show in Fig. 4a the average separations of nearest-neighbour iron-iron pairs, for both parallel and anti-parallel orientations of their local moments: $d_{FFP}$ and $d_{FFAP}$ respectively. It is clear that $d_{FFAP}$ is significantly smaller than $d_{FFP}$, i.e. that the iron-iron bond contracts when the spin configuration is changed from parallel to anti-parallel. Thus, the bigger $X_{FFAP}$ is, the smaller is the average distance between two nearest-neighbour iron atoms, $d_{FF}$.

Furthermore, as shown in Fig. 4b, there is a robust positive correlation between the average lattice constant, $a$, and $d_{FF}$. Thus, $a$ tends to decrease with increasing $X_{FFAP}$.

Finally, we make two comments. First, the physical origin of the discrepancy between $d_{FFAP}$ and $d_{FFP}$ in random face-centered-cubic $Fe_{65}Ni_{35}$ alloy has already been discussed[20]. It seems that this difference arises from the dependence of the exchange parameter on the distance between iron-iron nearest-neighbour atoms. Second, we can easily explain the deviation from Vegard's law shown, for example, by the low-temperature lattice constant of fcc $Fe_{65}Ni_{35}$ alloy[2, 6]. We obtain *ab initio* that the lattice constant of fully magnetised fcc $Fe_{1-x}Ni_x$ decreases linearly with increasing the Ni atomic concentration, $x$, between $x=0.3$ and $x=0.8$, in obedience to Vegard's law. However, according to our Ising model, the zero-temperature magnetic structure of fcc $Fe_{65}Ni_{35}$ alloy is not fully magnetised, but rather collinear ferrimagnetic: there are some anti-parallel iron-iron nearest neighbour moments even at zero temperature. Consequently, due to the mechanism discussed above, the lattice constant is smaller for the collinear ferrimagnetic alloy than for the fully magnetised alloy. The predicted relative deviation from Vegard's law is -0.37%, which is in fair agreement with the experimental result of -0.11%[6].

**Methods.**

To carry out point 1 of the procedure for $Fe_{1-x}Ni_x$ for $x=0.30$, 0.35, 0.45, 0.55, and 0.80, a mean-field Ising model of the Müller-Hesse type[10] is employed. However, there is no reason why an *ab initio* calculation[14] could not replace this approach. The input parameters of our Ising model for $Fe_{1-x}Ni_x$ are chosen as follows. The magnitude of the local magnetic moment at an iron site $M_F(x)$ and at a nickel site $M_N(x)$ is derived from first-principles calculations for ferromagnetic $Fe_{1-x}Ni_x$. For $x=0.35$, $M_F(x)$ and $M_N(x)$ are respectively set to 2.63 $\mu_B$ and 0.62 $\mu_B$. The exchange constants $J_{FF}(x)$ (between a nearest-neighbour iron-iron pair) and $J_{FN}(x)$ (between a nearest-neighbour iron-nickel pair) are tuned in such a way that the calculated zero-temperature magnetization and Curie temperature agree with experimentally measured properties[3, 21] within 10%. The exchange constant $J_{NN}(x)$ (between a nearest-neighbour nickel-nickel pair) is set to 40.55 meV, the value of $J_{NN}$ obtained at $x=1$. Note that $J_{FF}(x)$, $J_{FN}(x)$, and $J_{NN}(x)$ are the *only* experimentally determined parameters in our model.

*Ab initio* total energy calculations for point 2 of the procedure are performed within the framework of the *exact muffin-tin orbitals* (EMTO) theory using the full charge density (FCD) technique[22]. The problem of substitutional chemical disorder is treated within the coherent potential approximation (CPA)[23]. The integration over the irreducible part of the Brillouin zone is done over approximately 500 **k**-points distributed according to the Monkhorst-Pack scheme[24]. This is sufficient to ensure that the calculated lattice constants, bulk moduli and Grüneisen constants are converged with respect to the number of **k**-points within 5 mÅ, 100 kbar, and 0.1 respectively. All the calculated bulk moduli were found between 1.5 and 1.9 Mbar, and all the calculated Grüneisen constants lie between 1.5 and 1.8.

The results displayed in Figs. 4a and 4b are determined *ab initio* by means of the projector augmented-wave (PAW)[25] method as implemented in the Vienna *Ab initio* Simulation Package (VASP)[26]. We apply the "special quasirandom structures" (SQS)[27] scheme to random $Fe_{65}Ni_{35}$ and $Fe_{20}Ni_{80}$. For each Ni atomic concentration ($x=0.35$ and 0.80), we construct a 96-atom SQS[28]. More details about the calculations for $x=0.35$ can be found in Ref. 20.

Finally, note that the *ab initio* computations are performed within the generalized gradient approximation (GGA)[29, 30] for the exchange-correlation energy functional.

1. Guillaume, C. E. Recherches sur les aciers au nickel. Dilatations aux températures élevées; résistance électrique. *C. R. Acad. Sci.* **125**, 235–238 (1897).

2. Hayase, M., Shiga, M. & Nakamura, Y. Spontaneous volume magnetostriction and lattice constant of face-centred cubic Fe-Ni and Ni-Cu alloys. *J. Phys. Soc. Japan* **34**, 925–931 (1973).

3. Wassermann, E. F. Invar: moment-volume instabilities in transition metals and alloys. In K. H. J. Buschow and E. P. Wohlfarth (eds.), *Ferromagnetic Materials* (Elsevier, 1990).

4. Moruzzi, V. L., Janak, J. F. & Schwarz, K. Calculated thermal properties of metals. *Phys. Rev. B* **37**, 790–799 (1988).

5. Vegard, L. Die Konstitution der Mischkristalle und die Raumfüllung der Atome. *Z. Phys.* **5**, 17 (1921).

6. Acet, M., Zähres, H., Wassermann, E. F. & Pepperhoff, W. High-temperature moment-volume instability and anti-Invar of γ-Fe. *Phys. Rev. B* **49**, 6012–6017 (1994).

7. Ullrich, H. & Hesse, J. Hyperfine field vectors and hyperfine field distributions in FeNi Invar alloys. *J. Magn. Magn. Mater.* **45**, 315–327 (1984).

8. Brown, P. J., Neumann, K.-U. & Ziebeck, K. R. A. The temperature dependence of the magnetization distribution in $Fe_{65}Ni_{35}$ Invar: incompatibility of the two-state model. *J. Phys.: Cond. Matt.* **13**, 1563–1569 (2001).

9. Brown, P. J., Jassim, I. K., Neumann, K.-U. & Ziebeck, K. R. A. Neutron scattering from Invar alloys. *Physica B* **161**, 9–16 (1989).

10. Müller, J. B. & Hesse, J. A model for magnetic abnormalies of FeNi Invar alloys. *Z. Phys. B* **54**, 35–42 (1983).

11. Rancourt, D. G. & Dang, M.-Z. Relation between anomalous magnetovolume behavior and magnetic frustration in Invar alloys. *Phys. Rev. B* **54**, 12225–12231 (1996).

12. van Schilfgaarde, M., Abrikosov, I. A. & Johansson B. Origin of the Invar effect in iron-nickel alloys. *Nature* **400**, 46–49 (1999).

13. Crisan, V. *et al.* Magnetochemical origin for Invar anomalies in iron-nickel alloys. *Phys. Rev. B* **66**, 014416 (2002).

14. Ruban, A. V., Khmelevskyi, S., Mohn, P. & Johansson, B. Temperature-induced longitudinal spin fluctuations in Fe and Ni. *Phys. Rev. B* **75**, 054402 (2007).

15. Abd-Elmeguid, M. M., Hobuss, U., Micklitz, H., Huck, B. & Hesse, J. Nature of the magnetic ground state in Fe-Ni Invar alloys. *Phys. Rev. B* **35**, 4796–4800 (1987).

16. Wildes, A. R. & Cowlam, N. Does non-collinear ferromagnetism exist in Invar? *J. Magn. Magn. Mater.* **272–276**, 536–538 (2004).

17. Khmelevskyi, S. *et al.* Large negative magnetic contribution to the thermal expansion in iron-platinum alloys: quantitative theory of the Invar effect. *Phys. Rev. Lett.* **91**, 037201 (2003).

18. Jiang, X., Ice, G. E., Sparks, C. J., Robertson, L. & Zschack, P. Local atomic order and individual pair displacements of $Fe_{46.5}Ni_{53.5}$ and $Fe_{22.5}Ni_{77.5}$ from diffuse x-ray scattering studies. *Phys. Rev. B* **54**, 3211–3226 (1996).

19. Robertson, J. L. *et al.* Local atomic arrangements in $Fe_{63.2}Ni_{36.8}$ Invar from diffuse X-ray scattering measurements. *Phys. Rev. Lett.* **82**, 2911–2914 (1999).

20. Liot, F. & Abrikosov, I. A. Local magnetovolume effects in $Fe_{65}Ni_{35}$ alloys. Accepted for publication in *Phys. Rev. B*.

21. Bando, Y. The magnetization of face centered cubic iron-nickel alloys in the vicinity of Invar region. *J. Phys. Soc. Japan* **19**, 237 (1964).

22. Vitos, L. Total-energy method based on the exact muffin-tin orbitals theory. *Phys. Rev. B* **64**, 014107 (2001).

23. Vitos, L., Abrikosov, I. A. & Johansson, B. Anisotropic lattice distortions in random alloys from first-principles theory. *Phys. Rev. Lett.* **87**, 156401 (2001).

24. Monkhorst, H. J. & Pack, J. D. Special points for Brillouin-zone integrations. *Phys. Rev. B* **13**, 5188–5192 (1976).

25. Blöchl, P. E. Projector augmented-wave method. *Phys. Rev. B* **50**, 17953–17979 (1994).

26. Kresse, G. & Furthmüller, J. Efficient iterative schemes for ab initio total-energy calculations using a plane-wave basis set. *Phys. Rev. B* **54**, 11169–11186 (1996).

27. Zunger, A., Wei, S.-H., Ferreira, L. G. & Bernard, J. E. Special quasirandom structures. *Phys. Rev. Lett.* **65**, 353–356 (1990).

28. Abrikosov, I. A. et al. Competition between magnetic structures in the Fe rich fcc FeNi alloys. *Phys. Rev. B* **76**, 014434 (2007).

29. Perdew, J. P., Burke, K. & Ernzerhof, M. Generalized gradient approximation made simple. *Phys. Rev. Lett.* **77**, 3865–3868 (1996).

30. Wang, Y. & Perdew, J. P. Correlation hole of the spin-polarized electron gas, with exact small-wave-vector and high-density scaling. *Phys. Rev. B* **44**, 13298–13307 (1991).

**Acknowledgements** We thank I. A. Abrikosov for interesting discussions. F. L. is grateful to B. Alling, M. Ekholm and P. Steneteg for helping him with calculations. This work was supported by grants from the Swedish Research Council (VR), the Swedish Foundation for Strategic Research (SSF), the Göran Gustafsson Foundation for Research in Natural Sciences and Medicine, the EPSRC (UK) and the Scottish Universities Physics Alliance.

**Figure 1 Calculated lattice constant of $Fe_{65}Ni_{35}$ as a function of temperature.** Circles: lattice constant calculated assuming the magnetic configuration to be fixed (see key). Dotted lines: fits to circles. From top to bottom, $(x_{FU}, x_{NU})$ = (0.90, 1), (0.89, 1), (0.84, 1), (0.79, 0.97), (0.71, 0.89), (0.57, 0.64), and (0.5, 0.5). Crosses: lattice constant calculated including thermally induced magnetic disorder. Solid line: direct interpolation between crosses.

**Figure 2 Magnetic and lattice properties of $Fe_{1-x}Ni_x$ for several values of $x$.** **a**, the fraction of iron-iron nearest-neighbour pairs that are anti-parallel as a function of temperature, according to a mean-field Ising model. **b**, the equilibrium lattice constant of $Fe_{1-x}Ni_x$ in a PDLM state versus the fraction of iron-iron nearest-neighbour pairs that are anti-parallel, according to *ab initio* calculations performed by means of the EMTO method.

**Figure 3 Lattice constant as a function of temperature, and thermal expansion coefficient at room temperature, of $Fe_{1-x}Ni_x$ for two different values of $x$. a**, the lattice spacing of $Fe_{1-x}Ni_x$ at equilibrium as a function of temperature, according to our calculations (crosses) and experiments[2] (triangles). Solid lines are direct interpolations between the calculated points. **b**, the linear thermal expansion coefficient of $Fe_{1-x}Ni_x$ at room temperature as a function of the concentration of nickel, according to our calculations (circles)

and experiments[3] (triangles). Solid lines are direct interpolations between the points.

**Figure 4 Average iron-iron nearest-neighbour bond lengths and lattice constant in $Fe_{1-x}Ni_x$ at $x$=0.35 and 0.80. a**, comparison of the average distance between two iron nearest-neighbours with parallel (triangles up) and anti-parallel (triangles down) moments in $Fe_{1-x}Ni_x$, determined by PAW computations. The magnetic state is collinear ferrimagnetic, with the fraction of anti-parallel iron-iron nearest-neighbour pairs, $X_{FFAP}$, as in Fig. 2a. **b**, the lattice constant of $Fe_{1-x}Ni_x$ versus the average iron-iron nearest-neighbour bond length, $d_{FF}$, according to *ab initio* calculations carried out by means of the PAW method. The regression line shows the tendency of the lattice constant to increase with increasing $d_{FF}$.

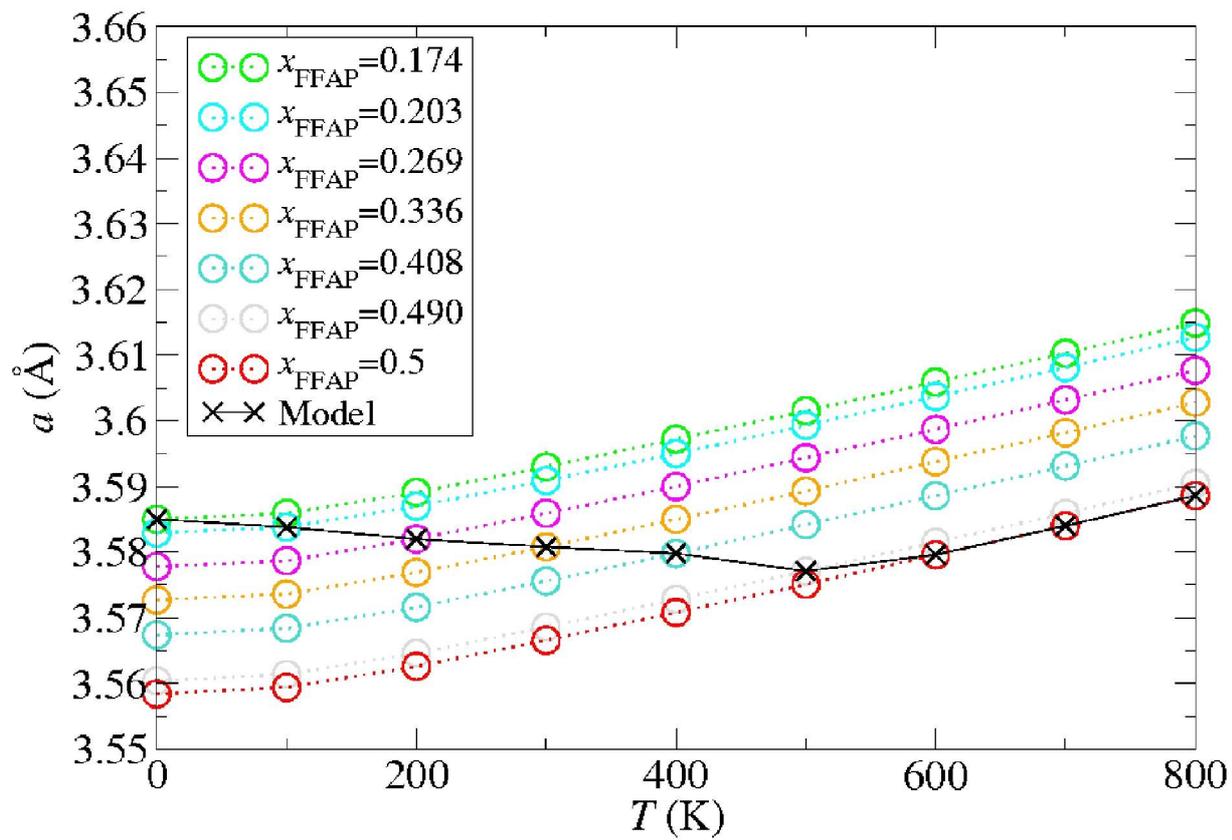

Figure 1.

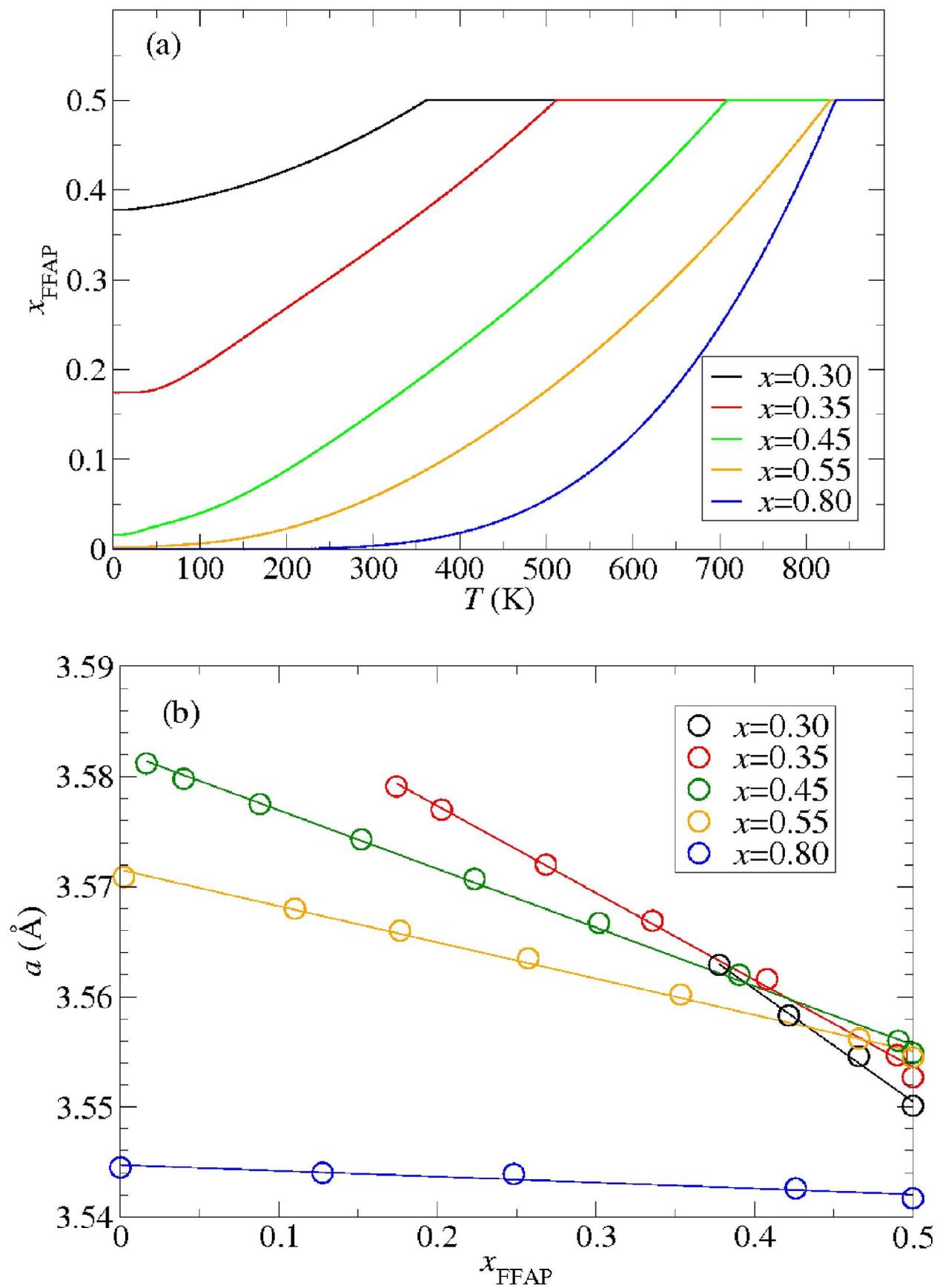

Figure 2.

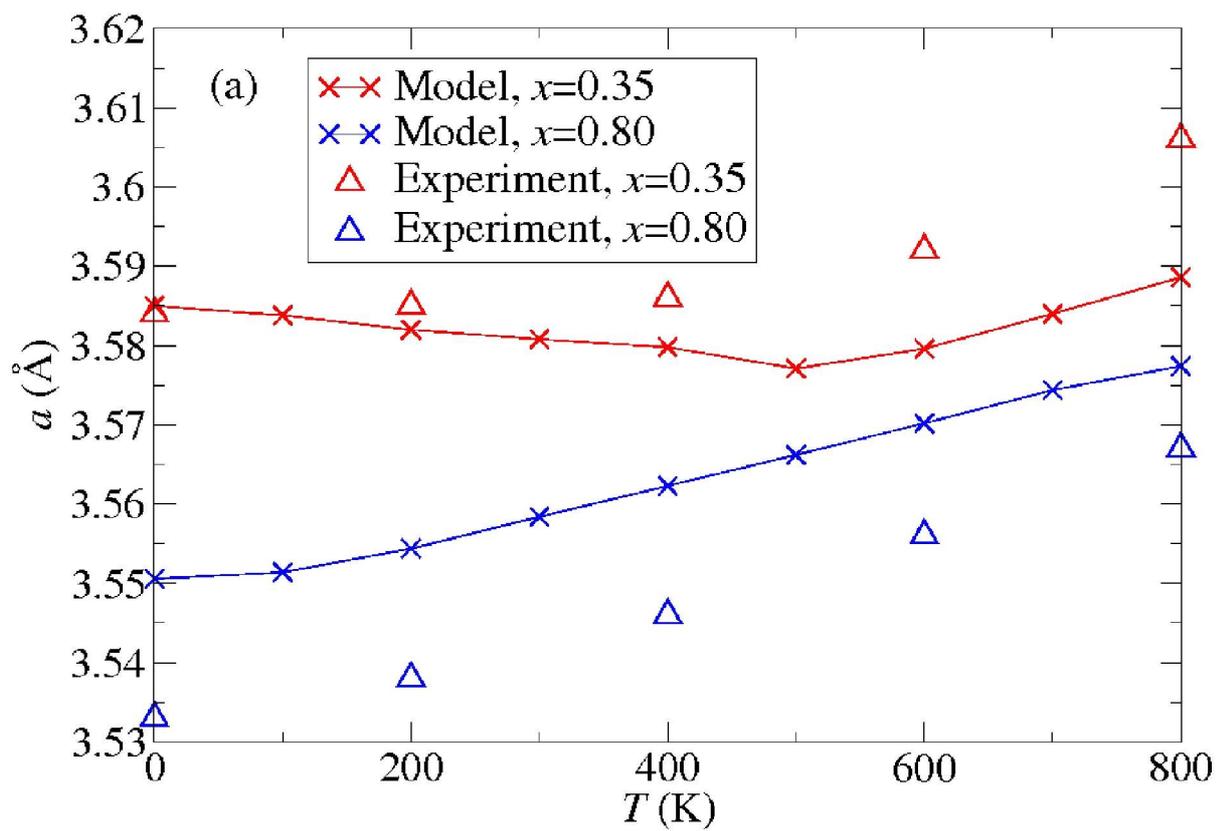
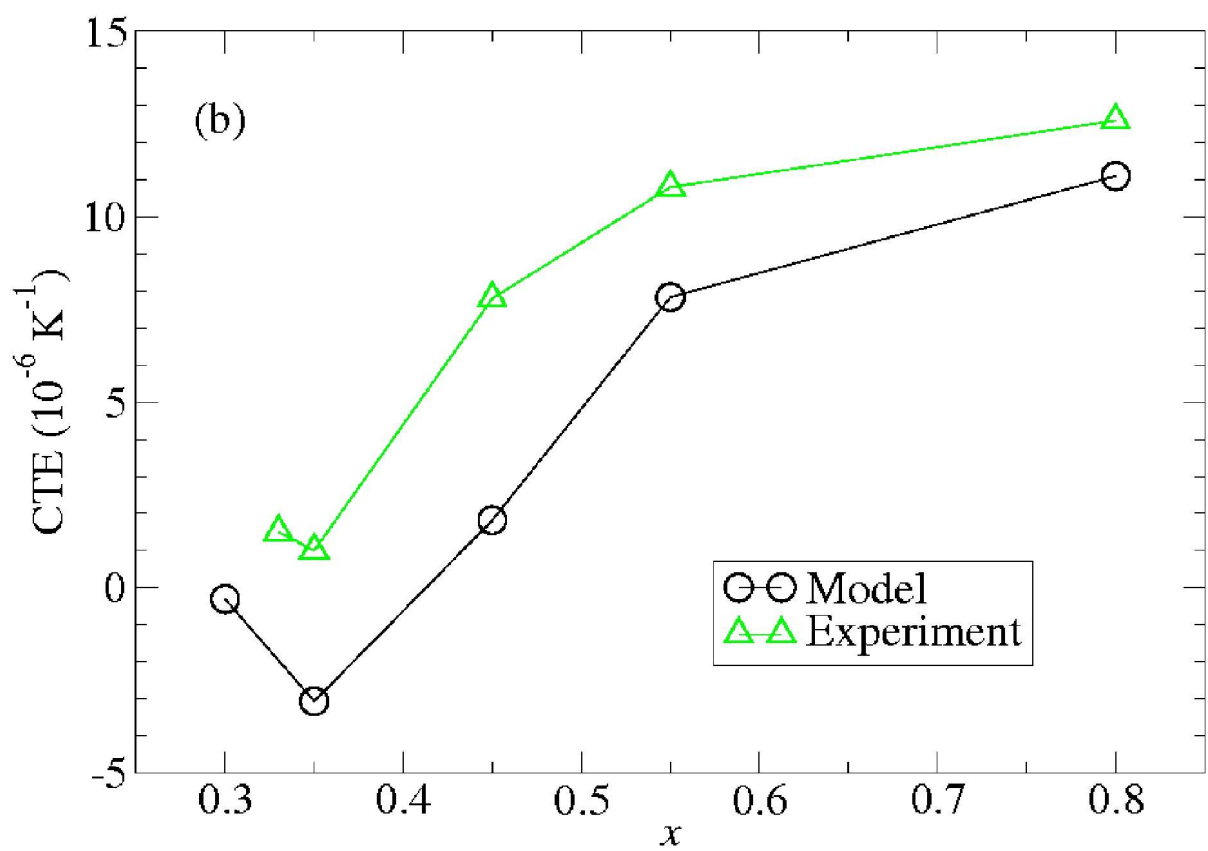

Figure 3.

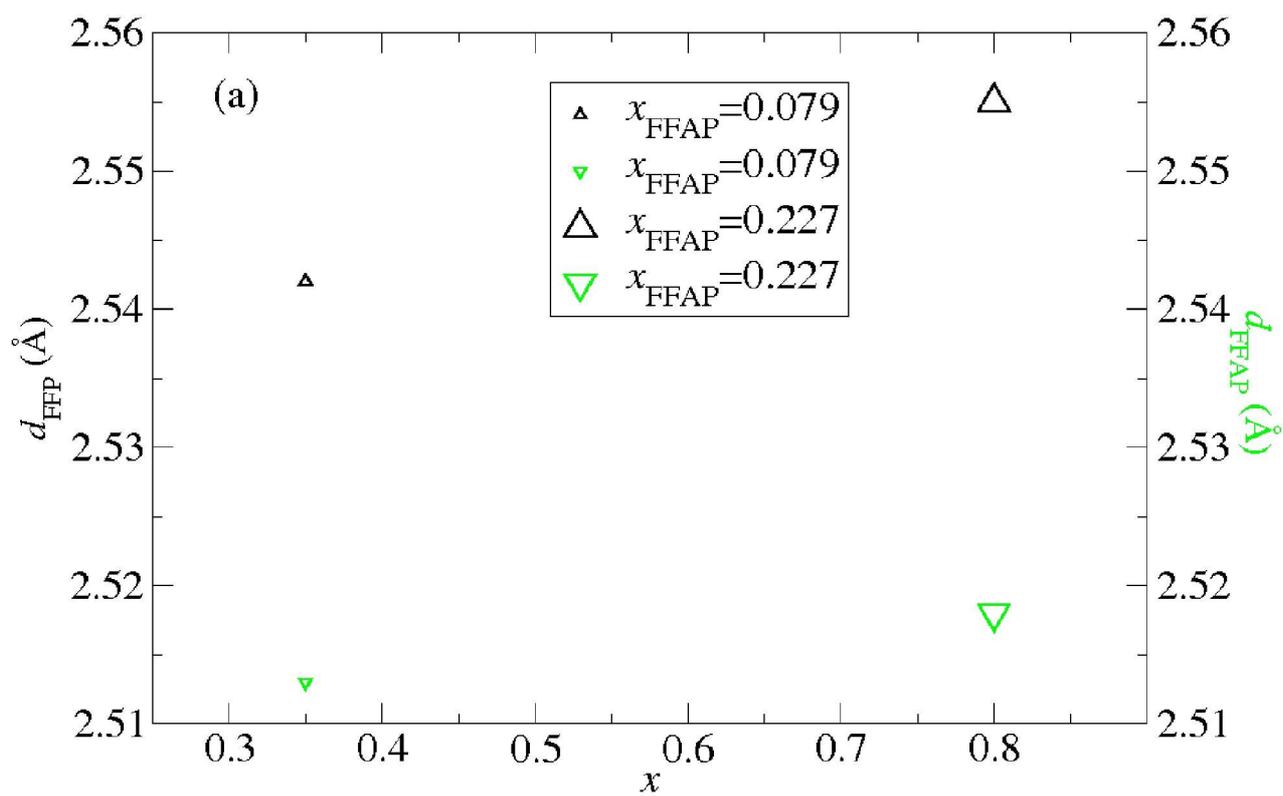
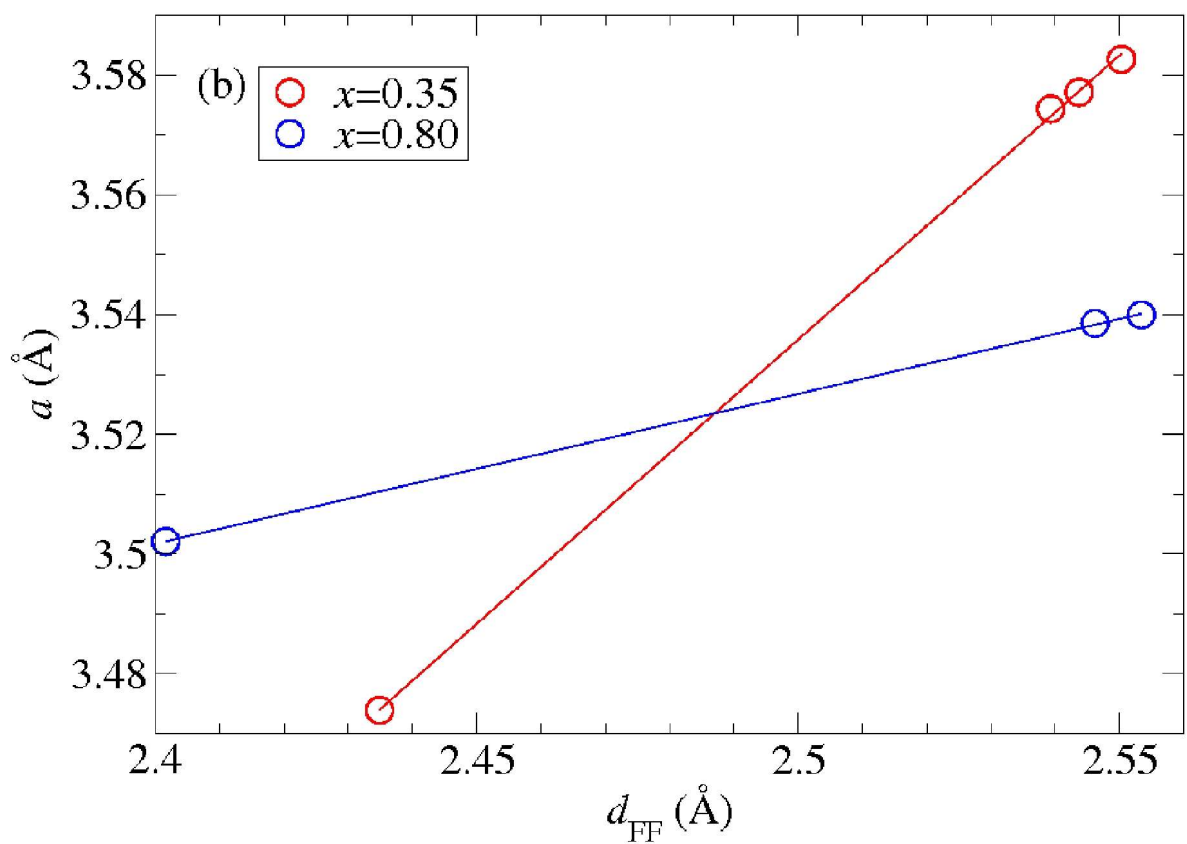

Figure 4.